\documentclass[aps,prb,twocolumn,showpacs,superscriptaddress,groupedaddress]{revtex4}  
\usepackage{graphicx}  % needed for figures
\usepackage{bm}        % for math
\usepackage{amsmath}
\usepackage{amssymb}   % for math
\usepackage{color}
\begin{document}
\title{Thickness Dependent Carrier Density at the Surface of SrTiO$_3$ (111) Slabs\\
}

\author{N.~Sivadas} 
\affiliation{Department of Physics, Carnegie Mellon University, Pittsburgh, Pennsylvania 15213, USA}
\author{H.~Dixit} 
\affiliation{Materials Science and Technology Division, Oak Ridge National Laboratory, Oak Ridge, Tennessee 37831, USA}

\author{Valentino~R.~Cooper} 
\email{coopervr@ornl.gov}
\affiliation{Materials Science and Technology Division, Oak Ridge National Laboratory, Oak Ridge, Tennessee 37831, USA}

\author{Di~Xiao}  
\email{dixiao@cmu.edu}
\affiliation{Department of Physics, Carnegie Mellon University, Pittsburgh, Pennsylvania 15213, USA}
\date{\today}

\begin{abstract}

We investigate the surface electronic structure and thermodynamic stability of the SrTiO$_3$  (111) slabs using density functional theory.  We observe that, for Ti-terminated slabs it is indeed possible to create a two-dimensional electron gas (2DEG).  However, the carrier density of the 2DEG displays a strong thickness dependence due to the competition between electronic reconstruction and polar distortions.  As expected, having a surface oxygen atom at the Ti termination can stabilize the system, eliminating any electronic reconstruction, thereby making the system insulating. An analysis of the surface thermodynamic stability suggests that the Ti terminated (111) surface should be experimentally realizable.  This surface may be useful for exploring the behavior of electrons in oxide (111) interfaces and may have implications for modern device applications. 
\end{abstract}
\maketitle

%%======== Introduction

\section{Introduction}

Since the discovery of a two-dimensional electron gas (2DEG) at the LaAlO$_3$/SrTiO$_3$ (LAO/STO) interface,~\cite{Ohtomo04p423} oxide heterostructures have attracted renewed attention.~\cite{hwang}  In addition to having high carrier densities~\cite{Seo07p266801} ($3\times 10^{14}$ /cm$^2$) with mobilities as high as $10^4$ cm$^2$/(V$\cdot$s),~\cite{Ohtomo04p423} such a 2DEG is also the host to a wide variety of electronic phenomena, including superconductivity,~\cite{Reyren07p1196} magnetism,~\cite{Brinkman07p493} and Rashba spin-orbit coupling.~\cite{BenShalom10p126802, King12p117602} These phenomena, together with the demonstrated tunability of interface conductivity through electric,~\cite{Thiel06p1942,Cen08p298} chemical and photosensitive means,~\cite{Meevasana11p114} offer promising potential for device applications with novel functionality.~\cite{review,Assmann13p78701}  As a first step, it is imperative to understand the nature and the origin of the 2DEG.  

In the LAO/STO system, it is now generally accepted that the \emph{intrinsic} 2DEG behavior is driven by the polarization discontinuity between the non-polar STO substrate and the polar LAO film.~\cite{Siemons07p196802, Ohtomo04p423, Nakagawa06p204, Pentcheva09p107602, Bristowe09p45425, Stengel09p241103, Bristowe11p081001, Stengel11p205432}  Along the [001] direction, the LAO film consists of alternating layers of (LaO)$^+$ and (AlO$_2$)$^-$, which leads to a divergent electrostatic potential, i.e., the so-called polar catastrophe.  Electronic reconstruction,~\cite{Okamoto04p630, Hamann06p195403, Okamoto06p056802, Cooper12p235109, Siemons07p196802, Takizawa11p245124, Cantoni12p3952} which is facilitated by the presence of transition metal ions, and polar distortions,~\cite{ Pentcheva09p107602,Wang09p165130, Hamann06p195403, Okamoto06p056802, Stengel11p136803} are the two main competing mechanisms that counter this divergent potential. While polar distortions screen the electrons at the interface or surface, thereby partially compensating for the polar catastrophe, electronic reconstruction cancels the divergent potential through a transfer of charge between the interface and the surface, with the excess charge partially occupying the Ti $3d$ states, giving rise to a 2DEG.  As such, a fundamental understanding of how these two effects compete with each other is important for tuning the properties of 2DEGs at oxide heterointerfaces.

\begin{figure*}
\includegraphics[width=1.3\columnwidth]{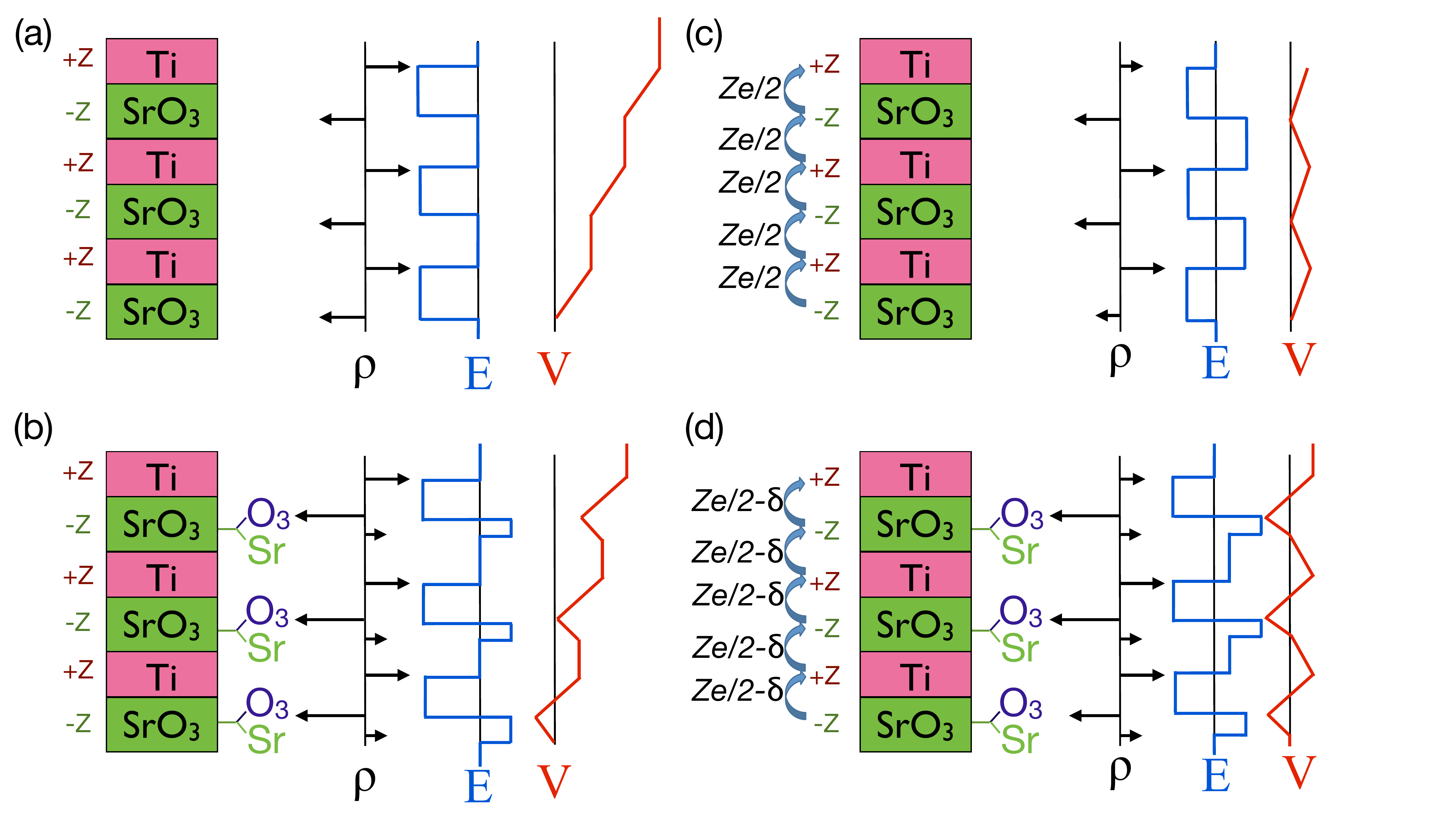}
\caption{\label{fig:unrelax} (Color online) Schematic of charge density ($\rho$), electric field ($E$), and electrostatic potential ($V$) profiles for the Ti-terminated STO (111) slabs with (a) charge uncompensated ideal atomic coordinates, (b) charge uncompensated relaxed atomic coordinates, (c) charge compensated ideal atomic coordinates and (d) charge compensated relaxed atomic coordinates. For the relaxed atomic coordinates (b) and (d), the SrO$_3$ layer (green block) splits into Sr (green line) and O$_3$ (violet line) layers. $Z$ is the magnitude of the effective charge of each layer.  Both polar distortions (b) and charge compensation due to electronic reconstruction (c) can help reduce the divergent electrostatic potential, and the net effect is less than $Ze/2$ electrons transferred between the surfaces (d).
}
\end{figure*}

Motivated by the above question, in this article we investigate the surface electronic structure of STO (111) slabs. Along the [111] direction, a STO slab consists of alternating stacks of nominally charged (Ti)$^{4+}$ and (SrO$_3$)$^{4-}$ layers, which already lead to a divergent electrostatic potential, even in the absence of a LAO overlayer (Fig.~\ref{fig:unrelax}).  This is in sharp contrast to a STO (001) slab, which consists of alternating charge neutral layers of SrO and TiO$_2$.  Furthermore, the presence of multivalent transition metal ions (Ti) suggests that electronic reconstruction can also take place in STO (111).  Hence, STO (111) slabs offer us a unique opportunity for studying both the polar distortion and electronic reconstruction in a chemically homogeneous system.  

Our work is also relevant to the recent experimental and theoretical progress on oxide (111) interfaces and surfaces.  Experimental studies have reported the growth of Ti-rich STO (111) surface~\cite{Biswas11p51904} and the creation of 2DEGs at (111) interfaces in the LAO/STO system, with carrier densities comparable to the [001] direction.~\cite{Herranz12p758, Annadi13p1838,eom}  Theoretically, exotic topological phases such as the quantum spin Hall state have also been predicted for cubic perovskite (111) bilayers.~\cite{Xiao11p596,Yang11p201104,fiete,satoshi} Understanding the surface electronic properties of the STO (111) slabs will have important implications for these phenomena as well.

Using density functional theory (DFT), we study STO (111) slabs of various thicknesses and different surface terminations.  We show that for Ti-terminated STO slabs (Figs.~\ref{fig:globfig}a and~\ref{fig:globfig}b) it is indeed possible to create a 2DEG.  However, the carrier density of the 2DEG displays a strong thickness dependence due to the competition between electronic reconstruction and polar distortion.  Our calculations suggest that relatively thick slabs ($\gg 12$ layers) are required to reach the ideal carrier density (2 electrons/surface unit cell) expected from the nominal charge counting argument.  In contrast, we find that the TiO-terminated slab (Figs.~\ref{fig:globfig}c and~\ref{fig:globfig}d) exhibits no charge transfer to the surface.  This is because the surface oxygen functions to nullify any potential that could develop across the STO slab. A thermodynamic stability analysis shows that the Ti-terminated STO slab can be stable, albeit within a very narrow region of phase space.  Our results show that both electronic reconstruction and polar distortions must be taken into account when analyzing the 2DEG behavior for (111) and (110) interfaces.~\cite{Herranz12p758}

%%======== Methodology

\section{Methodology}

\begin{figure}
\includegraphics[width=0.85\columnwidth]{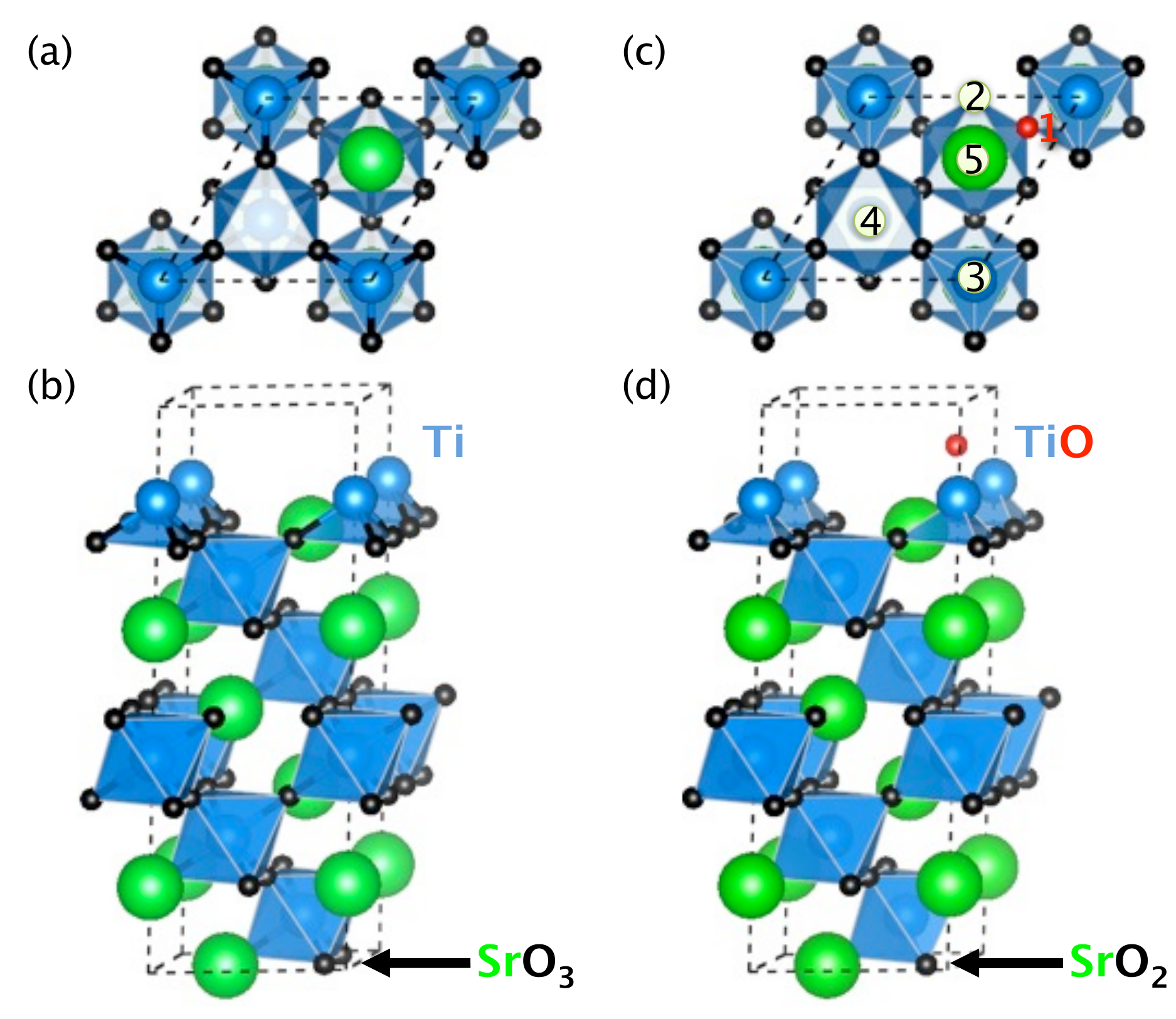}
\caption{\label{fig:globfig} (Color online)  Layer-by-layer structure of the $n$=6 STO (111) slab. (a) Top and (b) side view of the Ti-terminated STO (111) slab. (c) Top and (d) side view of the TiO-terminated STO (111) surface. Green, Blue and Black represent Sr, Ti and O ions, respectively. The five possibilities for placing the extra O atom on the Ti termination are also labeled in (c), and the energetically favorable position is marked by Red in both (c) and (d).}
\end{figure}

In this work we examine STO (111) slabs with various thickness and two surface terminations: Ti (top)/SrO$_3$ (bottom)-terminated (referred to as Ti-terminated) and TiO(top)/SrO$_2$(bottom) terminated (referred to as TiO-terminated).  Both are depicted in Fig.~\ref{fig:globfig}.

The electronic ground-state calculations were performed using DFT with the local density approximation (LDA) for exchange and correlation as implemented in the \textsc{quantum espresso} simulation package.~\cite{Giannozzi09p395502a}  We employ ultrasoft pseudopotentials\cite{Vanderbilt90p7892} including semicore electrons for O ($2s2p$), Sr ($4s4p5s$) and Ti ($3s3p4s3d$). To account for strong electronic correlations we use a Hubbard $U$ term (LDA+U).~\cite{Anisimov91p943}  For all calculations, a Hubbard $U$ = 5 eV for Ti $d$ states was found to be appropriate.  For each slab a 1$\times$1 in-plane periodicity with a vacuum region of $\sim$15 {\AA} was used.  A cutoff energy of 80 Ry and a Monkhorst-Pack special $k$-point mesh of 8$\times$8$\times$1 for the Brillouin zone integration was found to be sufficient to obtain less than 10 meV/atom convergence. We applied a dipole correction\cite{Bengtsson98p12301,Meyer01p205426} to set the electric field in the vacuum region to zero.  Structural optimizations were performed by fixing the in-plane lattice constant to that of the theoretical bulk STO lattice constant ($a_0$ = 3.85{\AA}).  All ions were then relaxed until the Hellmann-Feynman forces were less than 10 meV/{\AA}.

To analyze the thermodynamic stability of different surface terminations, we adopt a symmetric slab approach, i.e., the same termination on both sides. The thermodynamic analysis was performed using the generalized gradient approximations (GGA), as it usually gives more accurate formation energies, relative to experiments, than LDA.

%%======= Electronic Structure

\section{\label{sec:level3}Electronic structure}

\subsection{Ti-Termination}

\begin{figure}
\includegraphics[width=0.9\columnwidth]{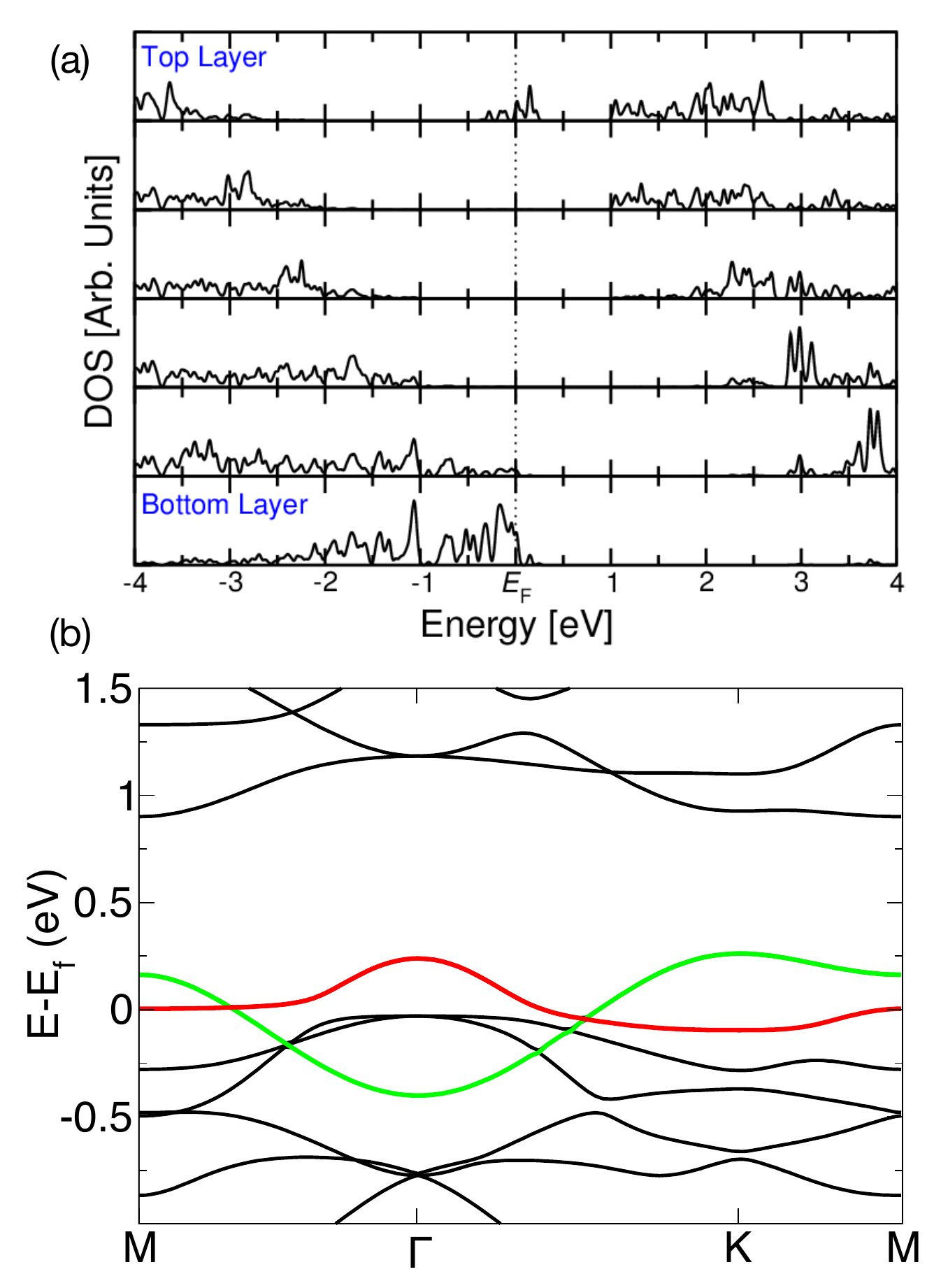}
\caption{\label{fig:stoichiometryLDOS} (Color online) (a) Layer-dependent DOS and (b) the electronic band structure for the Ti-terminated STO (111) slab with 6 layers.  There are two bands crossing the Fermi level; the top layer conduction band (green) consists mainly of Ti $d$ states, and the bottom layer valence band (red) consists mainly of O $p$ states.}
\end{figure}

We first consider the ideal Ti-terminated STO (111) slab, which consists of alternating stacks of Ti and SrO$_3$ layers with nominal charges of $+4$ and $-4$, respectively.  Figure~\ref{fig:stoichiometryLDOS}a shows the typical layer-projected DOS for a slab with 6 layers of SrTiO$_3$ unit.  We observe the occurrence of a surface metallic state for the top surface layer (Ti surface). This is similar to the results for the STO (110) surface\cite{Bottin03p35418} and with previous semi-empirical Hartree-Fock calculations for the STO (111) surface.\cite{Pojani99p177, Pojani99p179} The atomic projected DOS indicates that these surface states are comprised mainly of Ti $s$ and $d$ orbitals. The excess electrons are derived from the depletion of the valence band of the bottom layer due to the depopulation of O $p$ orbitals. This can be clearly seen from the orbital dependent electronic band structure, shown in Fig.~\ref{fig:stoichiometryLDOS}b.  
The dispersive energy bands clearly indicate that those excess charges are mobile carriers.

The number of excess electrons (holes) at the surface is obtained by performing an integration of the layer-averaged orbital projected DOS in a small neighborhood below (above) the Fermi level.~\cite{Stengel09p241103} (N.B. this typically underestimates the electron count relative to the total DOS by $\sim$0.1 e$^-$.) For the 6-layer slab, we found a total transfer of 0.7 electrons per surface unit cell.
However, if electronic reconstruction is the only working mechanism, nominal charge counting dictates that there should be two electrons per unit cell transferred from the SrO$_3$ surface to the Ti surface (Fig.~\ref{fig:unrelax}c).  This is a strong indication of the vital role  polar distortions~\cite{Wang09p165130, Hamann06p195403, Okamoto06p056802} play in avoiding the polar catastrophe in oxide heterostructures.~\cite{Pentcheva09p107602, Nakagawa06p204}.

\begin{figure}
\includegraphics[width=\columnwidth]{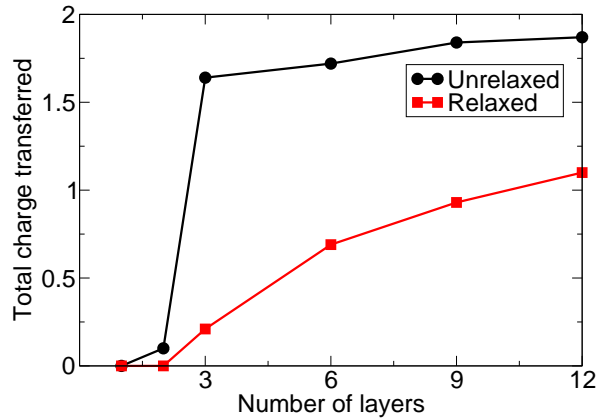}
\caption{\label{fig:totalchg} (Color online) The total charge transferred between the surfaces as a function of thickness for relaxed (red) and unrelaxed (black) system.}
\end{figure}

To analyze the effect of the polar distortions, we have calculated the total transferred charge between the surface layers for different layer thicknesses for both the relaxed and the unrelaxed systems (Fig.~\ref{fig:totalchg}).  There are two main features.  First, there is a critical thickness (3 layers) above which electronic reconstruction between the surfaces takes place, as indicated by the appearance of a 2DEG.  Second, we observe a smooth increase in transferred charge with a thickness dependence for the relaxed system, which is in sharp contrast to the abrupt increase of transferred charge with very little thickness dependence for the unrelaxed system.  We also note that for the unrelaxed structure, the transferred charge is very close to 2 electrons per unit cell.

The origin of the critical thickness can be explained by comparing this situation to the LAO/STO (001) heterostructure counterpart. It is well established that for these systems above an overlayer thickness of 3 layers, there is an occurrence of a 2DEG.\cite{Thiel06p1942} There the polar catastrophe is built up in the LAO overlayer and electronic reconstruction occurs when the divergent potential exceeds the band gap of STO. A similar effect occurs for the STO (111) slab.  When the slab is thick enough to make the divergence in potential comparable to the band gap an electronic reconstruction occurs.  This is clearly seen in our results for the unrelaxed surfaces (Fig.~\ref{fig:totalchg}).  In these systems, we observe a transfer of two electrons between the surfaces once the layer thickness is beyond a critical thickness of 2 layers. This is very abrupt and as immediate as the closing of the band gap and is a consequence of the fact that the electronic reconstruction is the only mechanism available for countering the polar catastrophe in the unrelaxed slabs.

%The thickness dependence of the transferred charge for the relaxed geometry can be accounted for by polar distortions.  

\begin{figure}
\includegraphics[width=\columnwidth]{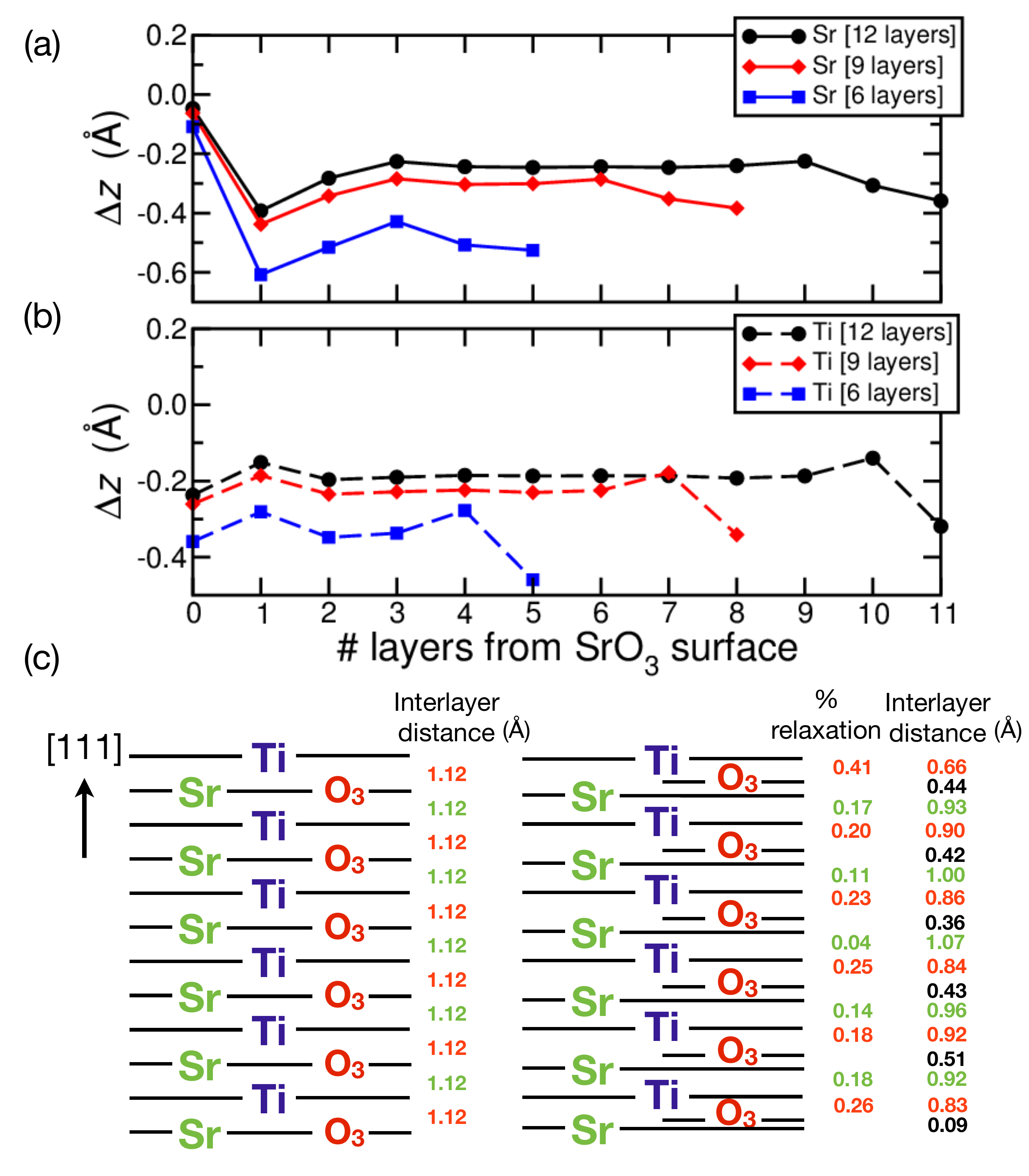}
\caption{\label{fig:pd1} (Color online) Polar distortions for the Ti-terminated STO (111) slab for layer thickness $n=$ 6, 9 and 12, relative to the SrO$_3$ bottom layer. (a) Split between Sr and O$_3$ in SrO$_3$ layers. (b) Off-centering of Ti relative to planes of O anions above and below Ti. Surface Ti off-centering is computed relative to bulk positions. (c) A schematic of the effect of relaxation for the case of Ti-termination with 6 layers of STO.}
\end{figure}

More interestingly, the thickness dependence of the magnitude of the charge transfer can be attributed to the effect of polar distortions. Figures~\ref{fig:pd1}a~and~\ref{fig:pd1}b depict the layer-by-layer off-center (polar) distortions, $\Delta z$, of Sr and Ti atoms, respectively, for Ti-terminated slabs of varying STO thicknesses. As can be seen from Fig.~\ref{fig:pd1}, the net effect of the (111) surface geometry and the Ti-termination is to polarize the individual layers of SrO$_3$ and Ti layers relative to their respective oxygen cages. For all slab thicknesses, we find that in the middle of the slab there is a nearly constant shift of Ti and Sr relative to their corresponding oxygen layers, with the magnitude of the displacements decreasing with slab thickness. More important are the large off-center displacements for the surface Ti cations. Figure~\ref{fig:pd1}c shows a schematic of the effect of relaxation for 6 layers of the Ti-terminated STO slab with the percentage relaxation calculated which emphasizes the large surface distortions (N.B. these large relaxations are similar to other oxides). 

The net effect of such surface dipoles is to counter the divergent potential across the slab. As the thickness of the slabs increase, however, we observe a decrease in the magnitude of these distortions. Below three layers the polar distortions are able to completely cancel out the polar catastrophe. As the thickness is increased, it becomes more energetically favorable to have an increase in the transferred charge and a decrease in polar distortion. We infer from the trend in Fig.~\ref{fig:totalchg} that for very thick STO slabs, the total amount of transferred electrons will converge to two electrons per unit cell after relaxation, just like in the case of the unrelaxed system. The competition between the electronic reconstruction and polar distortion is illustrated in Fig.~\ref{fig:unrelax}.

%One of the main distinctions between the unrelaxed and the relaxed system is that the critical thickness is 2 and 3 respectively. This and all the other thickness dependent features can be attributed to the effect of polar distortion. For the unrelaxed system
%The smooth vs.~abrupt behavior of the thickness dependent charge transfer is also an effect of polar distortion. 

%On the other hand, the smoother nature of the change in charge transfer with thickness for the relaxed systems can be explain by the gradual change in polar distortions with thickness. Because of this additional competing mechanism to cancel the polar catastrophe, the critical thickness now is a bit larger than the unrelaxed case. 

% In other words, the sum total of the large polar distortions is to reduce the overall charge transfer between the surfaces by screening the built-in potential.  This introduces a smoothness to the transferred charge as a function of thickness. This is the reason for the much smaller charge transfer compared with unrelaxed structures which is also consistent with the observation of an increase in the transferred charge with thickness as polar distortion decreases with thickness (see Figs.~\ref{fig:pd1}a~and~\ref{fig:pd1}b). 

\subsection{TiO-Termination}

Next we consider the TiO-termination of the STO (111) slab, which is made by transferring one O atom from the bottom SrO$_3$ terminated surface to the top Ti terminated surface, as shown in Figs.~\ref{fig:globfig}c and d. When we remove one O atom from the SrO$_3$ layer, the 6 identical O sites around the Sr atom are broken into a group of 4 identical sites and 2 identical O "vacant'' sites. To identify the most stable location for placing the O atom on the Ti termination, we compared the energies of 5 possible high symmetry sites as shown in Fig.~\ref{fig:globfig}c, and found that the O vacant site (1) is the preferred site.  

\begin{figure}

\includegraphics[width=\columnwidth]{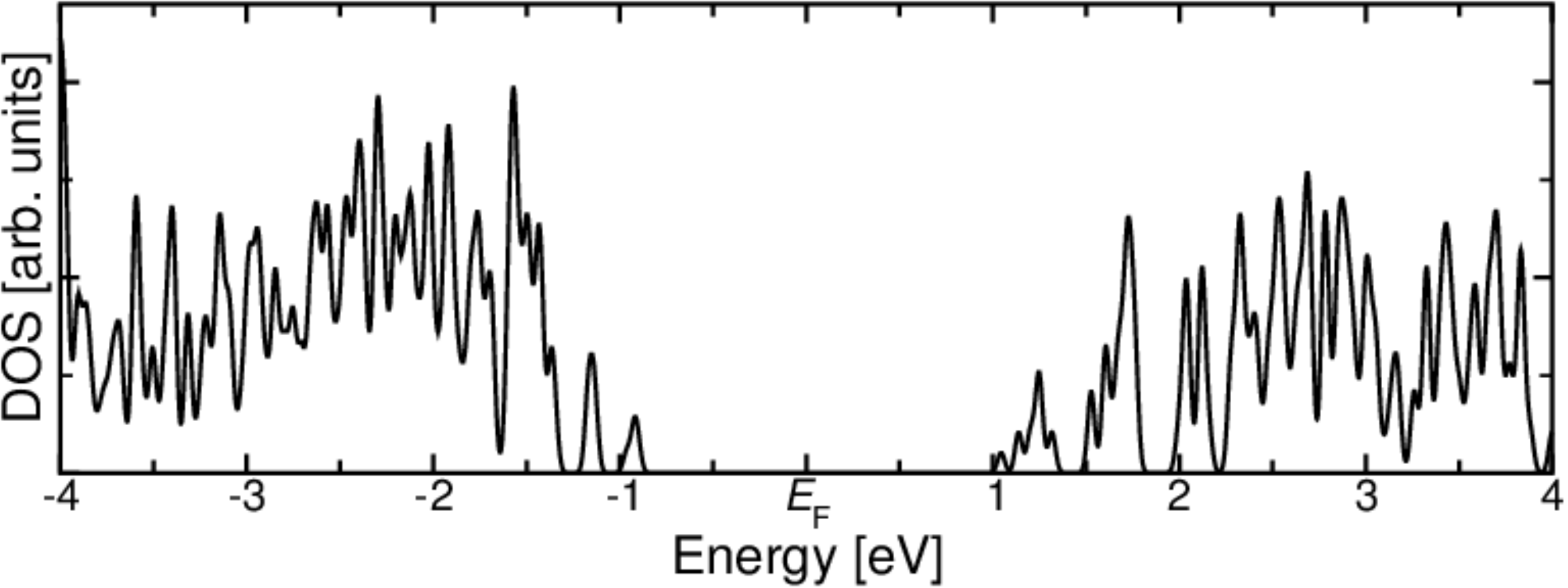}
\caption{The total DOS for 6 layers of the TiO-terminated STO (111) slab.}
\label{fig:odTDOS}
\end{figure}

We are interested in this situation because according to the polar catastrophe argument,  one should not expect any significant electron transfer between the surfaces as the transfer of one O atom between the surfaces should stabilize the system.  An analysis of the total DOS (see total DOS for 6 layers of STO in Fig.~\ref{fig:odTDOS}) confirms that the system is indeed insulating for all STO layer thicknesses studied ($n =$ 6, 9, 12).

\begin{figure}
\includegraphics[width=0.9\columnwidth]{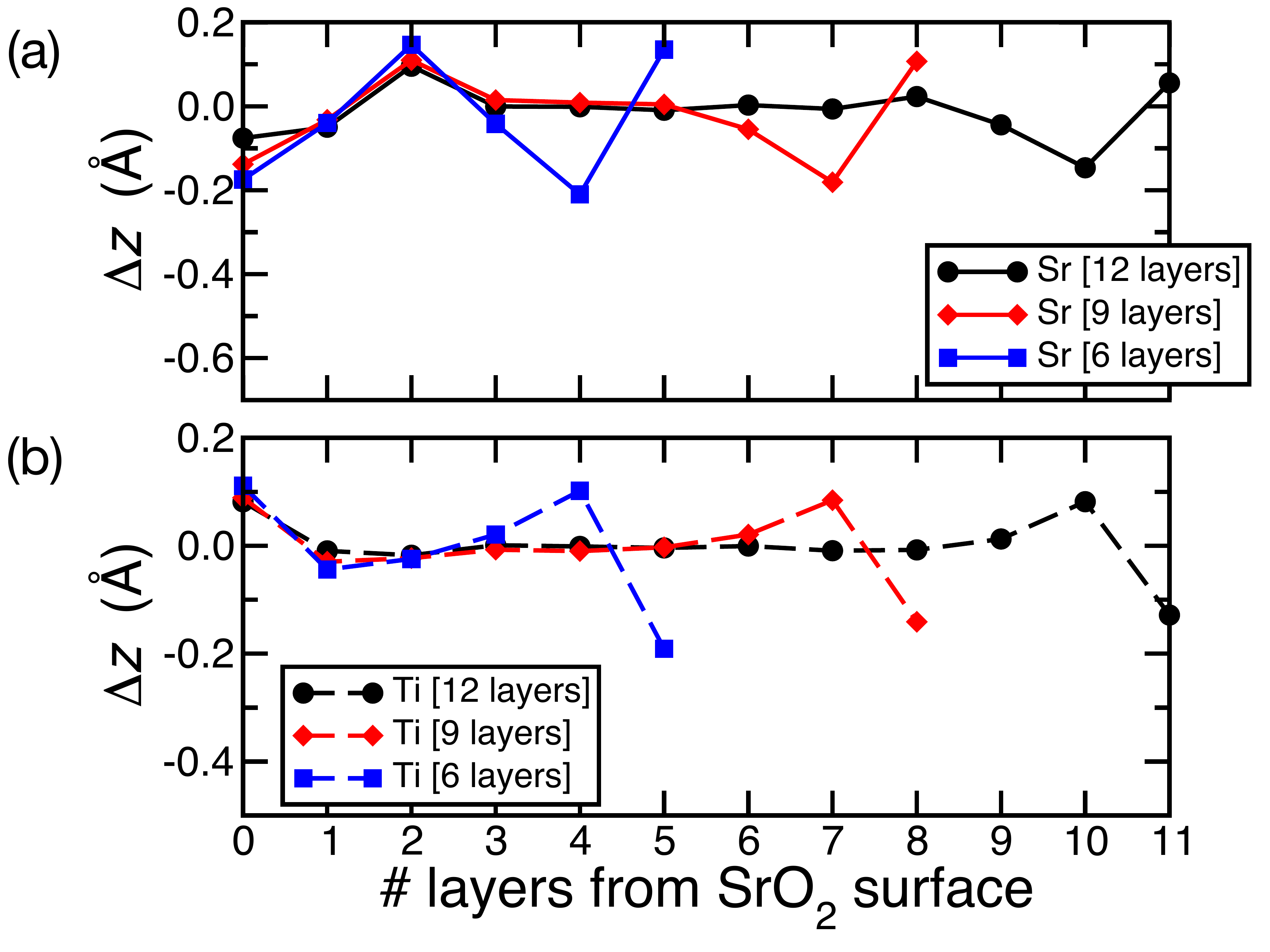}
\caption{\label{fig:pd2} (Color online) Polar distortions for the TiO-terminated STO (111) slab for layer thickness $n =$ 6, 9 and 12, relative to the SrO$_2$ bottom layer. (top) Split between Sr and O$_3$ in SrO$_3$ layers. (bottom) Off-centering of Ti relative to planes of O anions above and below Ti.}
\end{figure}

From this we note that there is no charge transfer between any bulk layers, similar to the Ti-terminated case. Nevertheless, we find a striking difference between the TiO- and Ti-terminated surfaces. In particular, in the case of the TiO-terminated surfaces there are no electrons transferred between the surface layers. Here, the polar catastrophe is avoided without any contribution from electronic reconstruction, unlike the case of the Ti-termination.

The atomic relaxation for each STO layer for different layer thickness of STO for the TiO-termination is shown in Fig.~\ref{fig:pd2}. Here we see that, unlike the Ti-terminated slabs, the bulk Ti and Sr cations exhibit negligible off-center displacements. Such displacements are indicative of having a reduced electric potential across the slab. However a larger polar distortion develops on the topmost plane of Ti ions, which is a local relaxation effect. To conclude, the effect of polar distortion in this case is much less compared to the Ti-termination  (Fig.~\ref{fig:pd1}) as the transfer of one oxygen atom is effective in screening the polarization, with the help of some local relaxation. This is in agreement with the observation of the lack of charge transfer for different thicknesses for this system.

\section{\label{sec:level4} Thermodynamic stability}

The imminent question is the relative thermodynamic stability of these different surface terminations. In this section we analyze the thermodynamic stability of the various terminations of the STO (111) surface, following the formalism proposed by F. Bottin \textit{et.al.}~\cite{Bottin03p35418} The energy required to split a crystal in half with complementary surfaces is called the cleavage energy. The cleavage energy (E$_{\rm cl}$) per surface area for the unrelaxed Ti/SrO$_{3}$ terminations and unrelaxed TiO/SrO$_{2}$ terminations respectively are defined in the following
\begin{equation}
\begin{aligned}
E_{\rm cl}^{Ti/SrO_3} = \frac{1}{2S}(E_{\rm slab}^{Ti} + E_{\rm slab}^{SrO_{3}} -nE_{\rm bulk}) \;, \\
E_{\rm cl}^{TiO/SrO_2} = \frac{1}{2S}(E_{\rm slab}^{TiO} + E_{\rm slab}^{SrO_{2}} -nE_{\rm bulk}) \;,	
\label{equ:cl}
\end{aligned}
\end{equation}

\noindent where E$_{\rm bulk}$ stands for the total energy of the bulk STO system, \emph{n} the total number of STO layers and $S$ denotes the surface area. Here E$_{\rm slab}^{\lambda}$ is the energy of ${\lambda}$ termination, with ${\lambda}$ being either Ti, SrO$_3$, TiO or SrO$_2$.

Since the cleavage energy does not distinguish between the two complementary surfaces, we define the surface energy, ${\Phi}_{\lambda}$, which is a measure of the stability of the surface with respect to bulk as:
\begin{equation}
\begin{aligned}
{\Phi}_{\rm \lambda} & = \frac{1}{2S}[E_{\rm slab}^{\lambda} -N_{Ti}E_{\rm bulk} -E_{O_2}^{\rm mol}/2(N_O-3N_{\rm Ti}) \\
 & -E_{Sr}^{\rm bulk}(N_{Sr}-N_{Ti})] \;.
\label{eq:phi}
\end{aligned}
\end{equation}

\noindent Table~\ref{tab:table3} shows the cleavage energy, relaxation energy and the surface energy for different terminations. The values for cleavage energy are in good agreement with previous first principles calculations for STO (111) surfaces.~\cite{Eglitis12p125}

\begin{table}

\caption{\label{tab:table3}The unrelaxed cleavage energy, relaxation energy and surface energy in J/m$^2$ for different terminations.}
\begin{ruledtabular}
\begin{tabular}{ccccc}
 &SrO$_3$ &Ti&SrO$_2$&TiO
 \\
\hline
E$_{cl}$& 6.62 &6.62  & 4.58 &4.58 \
\\
E$_{rel}$&-0.27 &-1.45  & -1.54 &-1.69 \
\\
${\Phi}_{\lambda}$ & 6.36 &  5.17& 3.04& 2.89\
\\
\end{tabular}
\end{ruledtabular}
\end{table}

However this definition of surface energy excludes the possibility of contact with matter reservoir. Hence we compute the surface grand potential,
\begin{equation}
{\Omega}_{\rm \lambda} = \frac{1}{2S}[E_{\rm slab}^{\lambda} -N_{Ti}{\mu}_{Ti} -N_{Sr}{\mu}_{Sr}-N_{O}{\mu}_{O}] \;,
\label{eq:omega}
\end{equation}
which is a measure of the excess energy of a symmetric system exposing a termination of a given composition, to a reservoir. 
The quantities $\mu_{Ti}$, $\mu_{Sr}$, ${\mu}_{O}$ in Eq.~(\ref{eq:omega}) are the chemical potentials of the Ti, Sr and O atomic species, respectively, and $N_{Ti}$, $N_{Sr}$, $N_{O}$ are the number of Ti, Sr and O atoms in the slab. 
The chemical potential of the bulk STO system (${\mu}_{SrTiO_3}$) can be written as the sum of chemical potentials of individual species in the crystal:
\begin{equation}
{\mu}_{SrTiO_3}= {\mu}_{Sr}+{\mu}_{Ti}+3{\mu}_{O}  \;.
\label{eq:mu}
\end{equation}
When the surface is in equilibrium with the bulk, we have ${\mu}_{SrTiO_3}$ = E$_{bulk}$. Using this and Eq.~(\ref{eq:mu}) we can rewrite the surface grand potential in Eq.~(\ref{eq:omega}) as:  
\begin{equation}
\begin{aligned}
{\Omega}_{\rm \lambda} & = \frac{1}{2S}[E^{\lambda}_{\rm slab} -N_{Ti}E_{\rm bulk} - {\mu}_{O}(N_{O}-3N_{Ti}) \\
& - {\mu}_{Sr}(N_{Sr}-N_{Ti})  ] \;.
\label{eq:omegainter}
\end{aligned}
\end{equation}
From this we observe that the range of accessible values for the surface grand potential depends on the maximum and minimum values of $\mu_{\rm Sr}$ and $\mu_{\rm O}$ chemical potentials. The possible variations in $\mu$ reflect the experimental growth conditions. Under the O rich condition $\mu_O=E_{O_2}^{\rm molecule}/2$  and for the Sr rich condition $\mu_{Sr}=E_{Sr}^{\rm bulk}$. Defining $\Delta\mu_O=\mu_O^{SrTiO_{3}}-E_{O_2}^{\rm molecule}/2$ and $\Delta\mu_{Sr}=\mu_{Sr}^{SrTiO_{3}}-E_{Sr}^{\rm bulk}$ we obtain 

\begin{equation}
\begin{aligned}
{\Omega}_{\lambda} = {\Phi}_{\lambda}-\frac{1}{2S}[{\Delta}{\mu}_{O}(N_O-3N_{Ti}) +{\Delta}{\mu}_{Sr}(N_{Sr}-N_{Ti})  ] \;.
\label{eq:omegamu}
\end{aligned}
\end{equation}

The ranges of the two independent parameters, ${\Delta}{\mu}_{O}$ and ${\Delta}{\mu}_{Sr}$ can be determined using the following set of conditions. In order to avoid the elements precipitating into Sr bulk, Ti bulk and oxygen gas, the upper bounds are set by 
$ \Delta \mu_{Sr} $, $ \Delta \mu_{Ti} $ \& $\Delta \mu_{O} \le 0 $. The lower bounds are obtained using:
\begin{equation}
\Delta \mu_{Sr}+3\Delta \mu_{O} > -E^f_{SrTiO_{3}} \;,
\end{equation}
where the formation energy ($E^f_{SrTiO_{3}}$) is defined as
\begin{equation}
-E^f_{SrTiO_{3}} = E^{\rm bulk}_{SrTiO_{3}} - E^{\rm bulk}_{Ti} - E^{\rm bulk}_{Sr} - \frac{3}{2}E^{\rm molecule}_{O_2} \;.
\label{eq:omegamu3}
\end{equation}

\begin{figure}
\includegraphics[width=0.9\columnwidth]{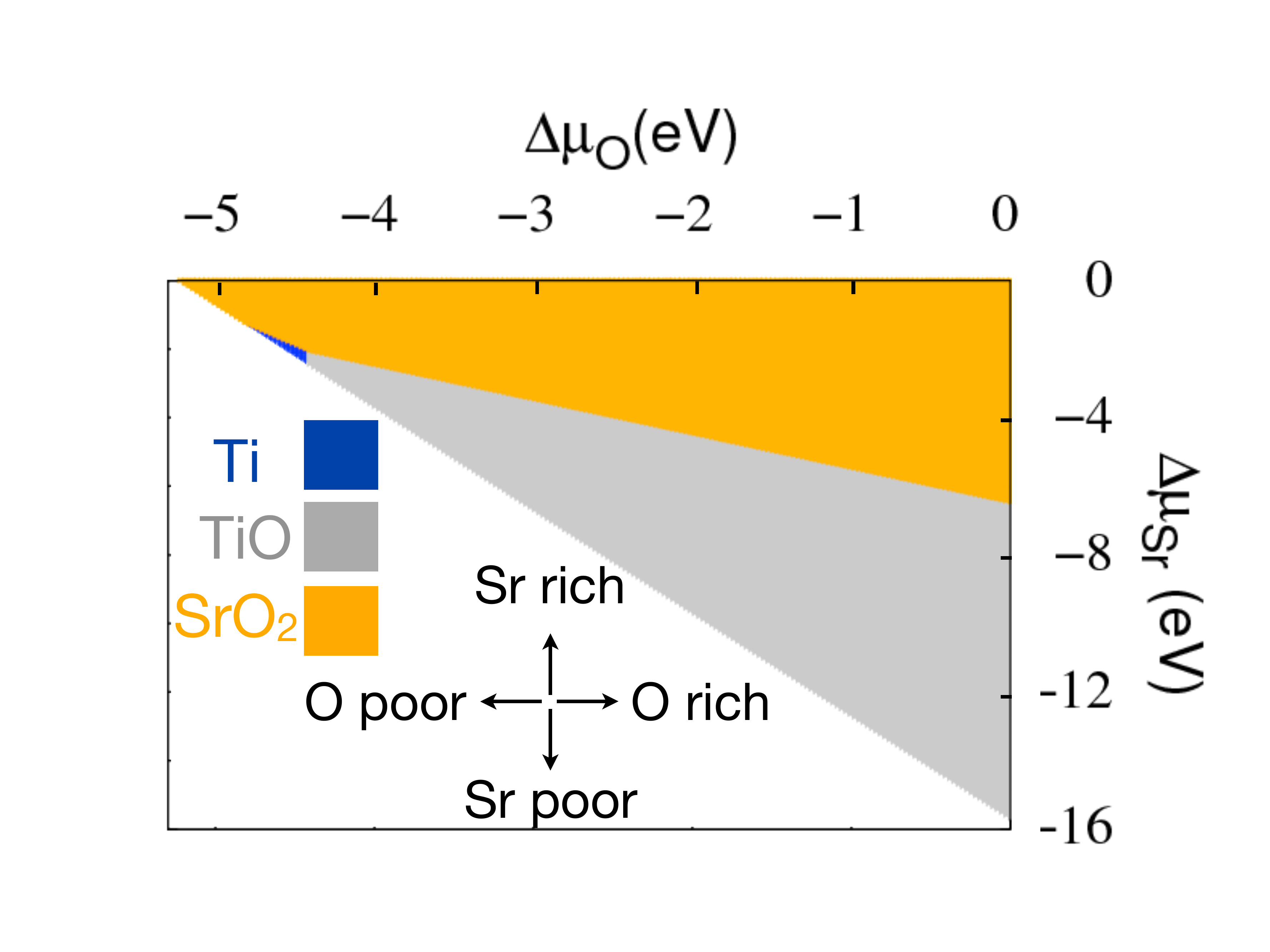}
\caption{\small{ (Color online) Stability diagram of the STO (111) surface. The most stable termination is shown as a function of the chemical potential of O and Sr.}}
\label{fig:stab}
\end{figure}

Figure~\ref{fig:stab} shows the relative stability of different terminations in the (${\Delta}{\mu}_{O}$,${\Delta}{\mu}_{Sr}$) plane after computing $\Omega$. The shaded area shows the region in the chemical potential phase space where a particular termination is most stable. We observe that the SrO$_2$-SrO$_2$ termination is most stable under O and Sr rich condition. When $\mu_{\rm Sr}$ is lowered, the TiO-TiO termination becomes the most stable surface. However for low oxygen pressure, there is also a region where the Ti termination is stable, which implies that it is possible to observe the effect of electronic reconstruction by tuning the experimental conditions for STO (111) surface. It should be noted that for the STO (111) surface a stable Ti-rich surface has already been observed (albeit without any knowledge of O content).\cite{Biswas11p51904} Nevertheless, these results highlight the possibility of creating the Ti-terminated surface and may have specific consequences for the electronic states that could be created at a heterostructure interface.

\section{Conclusion}

In summary, we have studied STO (111) slabs of various thicknesses and different surface terminations using density functional theory. We observe that for the Ti-SrO$_3$ terminated STO (111) slab there is charge redistribution which is dominated by the transfer of electrons from the SrO$_3$ terminated surface to the Ti surface, giving rise to a metallic surface states for this configuration.  The carrier density of the 2DEGs display a strong thickness dependence due to the competition between electronic reconstruction and polar distortions.  In comparison, for the TiO-terminated surface, no such surface states exist and the compensation mechanism is dominated by the new surface boundary conditions created by the transfer of an O ion from one surface to the other. By studying the relative stability of these different terminations we observe that the Ti termination can indeed be stabilized depending upon the experimental conditions. Naturally, the ability to tune the magnitude of charge transfer/compensation at an oxide surface/interface has consequences on numerous applications including surface catalysis and oxide electronics as well as important implications for novel phenomena such superconductivity and magnetism in confined dimensions.

\begin{acknowledgements}
We would like to thank Marek Skowronski, Michael Widom, Satoshi Okamoto and Wenguang Zhu for useful discussions. This work was supported by AFOSR Grant No.~FA9550-12-1-0479 (N.S. and D.X.), and by the U.S. Department of Energy, Office of Basic Energy Sciences, Materials Sciences and Engineering Division (D.X., H.D. and V.R.C.) and the Office of Science Early Career Research Program (V.R.C). N.S. acknowledges support through the ASTRO program at ORNL. This research used resources of the National Energy Research Scientific Computing Center, supported by the Office of Science, US Department of Energy under Contract No.~DEAC02-05CH11231, and Pittsburgh Supercomputing Center's Blacklight system, supported by NSF Grant No.~OCI-1041726, which is part of the Extreme Science and Engineering Discovery Environment (XSEDE), supported by NSF Grant No.~OCI-1053575.
\end{acknowledgements}

\end{document}